\documentstyle[aaspptwo]{article}

\lefthead{Morris {\it et al.}}
\righthead{Lyman $\alpha$ Absorbers}

\begin{document}

\title{The Environment of Lyman $\alpha$ Absorbers\\ in the Sightline towards
3C273}

\author{S. L. Morris\protect\footnote{Present Address: Institute of
Astronomy, Madingley Rd., Cambridge CB3 0HA, UK}, R. J. Weymann\\ Alan
Dressler and P. J. McCarthy}
\affil{The Observatories of the Carnegie Institution of Washington,\\ 813 Santa
Barbara St., Pasadena, CA 91101}

\author{B. A. Smith}
\affil{Institute for Astronomy, University of Hawaii, Honolulu, HI 96822}

\author{R. J. Terrile}
\affil{Jet Propulsion Laboratory, Pasadena, CA 91109}

\author{R. Giovanelli}
\affil{Department of Astronomy and National Astronomy and Ionosphere
Center\protect\footnote{The National Astronomy and Ionosphere Center is
operated by Cornell University under a cooperative agreement with the
U.S. National Science Foundation}, Cornell University, Ithaca, NY
14853}

\and
\author{M. Irwin}
\affil{Royal Greenwich Observatory, Madingley Rd, Cambridge CB3 0EZ, UK}

\begin{abstract}

We present new ground-based data following up on the HST discovery of
low-redshift Lyman $\alpha$ absorption in the sight-line to the quasar
3C273. Our goal is to investigate the relationship between the
low-column-density absorbers and higher column-density objects such as
galaxies or H~II regions. Narrow-band filter observations with a
coronograph show that there are no H~II regions or other strong H$\alpha$
line-emitting gas within a 12 kpc radius of the line-of-sight to the
quasar, at the velocities of three of the absorbers. Broad-band imaging
in Gunn r shows that there are no dwarf galaxies at Virgo distances
with absolute magnitude above M$_{\rm B}\approx$-13.5 and within a
radius of 40 kpc from the line-of-sight to the quasar.  Finally, we
present fiber spectroscopy of a complete sample of galaxies within a
radius of 1{\deg}, down to an apparent magnitude of B$\approx$19.
Analysis of this sample, combined with galaxies within 10 Mpc of the
quasar line-of-sight taken from the literature, shows that the
absorbers are definitely not distributed at random with respect to the
galaxies, but also that the absorber-galaxy correlation function is not
as strong as the galaxy-galaxy correlation function on large scales. We
show that our data are consistent with the hypothesis that all galaxies
more luminous than 1/10 L$^*$ have effective cross-sections (for
association with absorbers whose neutral-hydrogen column-density
(Log(NH)) is $>$13.0), of between 0.5 and 1 Mpc.  We also show a clear
case of a Lyman $\alpha$ absorber which has no galaxy brighter than
M$_{\rm B}$=-18 within a projected distance of 4.8 Mpc, and discuss the
possibility that Lyman $\alpha$ absorbers are destroyed in a rich
galaxy environment.

\end{abstract}


\section{Introduction} \label{introduction}

Understanding the origin and evolution of structure in the universe
remains one of the most fundamental and active challenges of current
astrophysical research. As the evidence in favor of a cosmological
origin for the narrow, displaced absorption lines in QSO spectra became
overwhelming, it also became clear that both the metal-line systems and the
Lyman $\alpha$ systems are invaluable tools for the study of some aspects of
this problem. Since ground-based Lyman $\alpha$ studies refer only to
redshifts $\ga$1.6, they complement studies of galaxy clustering
properties, the majority of which involve redshifts much less than
this.  However, precisely because the redshift regimes have been so
different and because it has not been at all clear what relation exists
between the typical low-column-density Lyman $\alpha$ absorbers and
galaxies, these two approaches have remained disjoint. It was somewhat
unexpected, but pleasing, that low-redshift Lyman $\alpha$ absorbers were found
in
sufficient numbers to enable meaningful  studies of the evolution of
the Lyman $\alpha$ absorbers and their relation to galaxies
(\cite{mor91,bah91b}). This has presented astronomers with the
opportunity to join these two lines of investigation.

There are two levels at which such such attempts can be carried out: 1)
Purely statistical investigations aimed at comparing the clustering
properties of galaxies and Lyman $\alpha$ absorbers, and 2)
Investigation of individual cases in which the possibility of
establishing the presence (or absence) of a clear link between a Lyman
$\alpha$ absorption line and something we could call a ``galaxy''
presents itself.  Preliminary discussions along these lines may be
found in papers by \cite{bah92a,bah92b} and by \cite{sal92}. The
present paper is a first attempt to pursue both these approaches along
the sightline to 3C273. The remainder of this paper is organized as
follows: In \S~\ref{observations} we describe the different sets of
observations we have assembled to investigate the environment of the Lyman
$\alpha$ absorbers along the 3C273 sightline. In \S~\ref{analysis} we
analyze them for possible associations or lack of associations of
individual Lyman $\alpha$ absorbers  with galaxies, and also give some
statistical analysis of the clustering properties of the Lyman $\alpha$
absorbers with galaxies. In \S~\ref{sumdisc} we discuss these results
in light of current models for the Lyman $\alpha$ absorbers and provide
a brief summary and suggestions for further work.

Throughout this paper H$_0$ is taken to be 80 km/s/Mpc, the distance to
the Virgo cluster is taken to be 16.0 Mpc (a distance modulus of
(m-M)$_{\rm Virgo}$=31.02) (\cite{jac92}), and q$_0$ is taken to be 0.

\section{Observations and Reduction} \label{observations}

It has long been realized that imaging of the gas directly responsible
for the low-column-density Lyman $\alpha$ absorbers is well beyond the
reach of current technology. Specifically, the neutral-hydrogen column
densities of order $10^{13}-10^{14}$ cm$^{-2}$ that GHRS detected
toward 3C273 are about 4 or 5 orders of magnitude below what can be
imaged in 21 cm emission, even without taking into account the powerful
radio background contributed by 3C273 itself. The H$\alpha$
recombination surface brightness associated with these neutral-hydrogen
column-densities is also several orders of magnitude below what is
feasible to detect, unless the incident flux of ionizing photons is
several orders of magnitude higher than that expected from the
integrated background radiation.

However, it has frequently be suggested that the Lyman $\alpha$
absorbers are intimately connected with, or are actual extensions of,
entities which {\it can} be imaged either by means of H$\alpha$
emission, starlight, or 21 cm emission - e.g. dwarf galaxies
(\cite{fra82}) or shells of expanding gas (\cite{che83}) or the outer
regions of galactic disks (\cite{mal92}). In the case of dwarf
irregulars, for example, a very small episode of recent star formation
might betray the presence of a dwarf irregular whose outer envelope
produces the Lyman $\alpha$ absorbers.  Alternatively, expanding shells
of gas might produce H$\alpha$ emission via collisional ionization at a
shock front.

In addition, of course, possible association of individual Lyman
$\alpha$ absorbers with specific galaxies, as well as statistical
studies of absorber-galaxy correlation can be carried out with a sample
of redshifts for galaxies in the field surrounding 3C273.

In this section we describe three such new sets of observations of a
region centered on 3C273. These are: \S~\ref{obscoron} Coronagraph
observations with narrow-band filters of a 5{\arcmin} diameter region,
\S~\ref{obscosmic} Deep broad-band imaging of a 17{\arcmin} diameter
region, and \S~\ref{obsfiber} Fiber spectroscopy of a
2.2$\times$1.6{\deg} region down to a limiting magnitude of B=19.  We
describe the analysis of these three sets of observations in
\S~\ref{analysis}

\subsection{Coronagraph Observations with Narrow Band Filters} \label{obscoron}

Observations of a 5\farcm3$\times$5\farcm3 region (radius $\approx$12
kpc at Virgo) around 3C273 were obtained during 1992 February 3-7, with
the University of Hawaii Coronagraph (\cite{vil87}) on the Las Campanas
2.5m duPont Telescope. A thinned 1024$\times$1024 Tektronics CCD was
used, binned 2$\times$2, giving a scale of 1.23\arcsec/pixel. The
coronagraph blocking mask had a diameter of 5\arcsec. Data were
obtained with a Gunn r filter, and also 5 specially acquired filters, 3
with width 13.5{\AA}, centered at 6586.2{\AA}, 6598.0{\AA} and
6756.0{\AA} (hereafter referred to as VN1, VN2 and HN), and 2 with
width 25{\AA}, centered at 6643.2{\AA} and 6718.9{\AA} (hereafter
referred to as VB and HB). (The above widths and central wavelengths
are quoted for an f7.5 beam and a temperature of 15{\deg} C). The
narrow-band wavelengths were chosen to match the redshifted position of
H$\alpha$ at the velocities of the 2 Lyman $\alpha$ absorbers listed in
\cite{mor91} at velocities corresponding to the Virgo cluster
(\cite{bin85}), and one absorber at 1251{\AA}.  This is the lowest
redshift, strong Lyman $\alpha$ system beyond the Virgo cluster.  The
observing procedure involved cycling through the 6 different filters,
with exposure times of 10 mins for the r and 25{\AA} filters, and 20
mins for the 13.5{\AA} filters. Seven such cycles were completed over
the 5 night run. During the observing run, it was discovered that the
narrow-band filter HN had slightly non-parallel faces, resulting in
detectable `ghost' images offset from bright stars and also 3C273. In
an attempt to minimise the effect of these, this filter was rotated
though 90{\deg} between each night.

For calibration purposes, observations were also obtained of Mkn 49 (an
emission line galaxy in the Virgo cluster with radial velocity 1524
km/s, and hence with H$\alpha$ line-emission within 2{\AA} of the peak
of VN2) and M87, and also a number of bright standard stars.

The images were reduced using IRAF\footnote{IRAF is distributed by
NOAO, which is operated by AURA Inc. under contract to the NSF}. The
reduction steps were: bias subtraction, division by a flat field taken
on the same night as the observations, rotation and shifting of the
images to match a reference image, sky subtraction and averaging
together of images taken with the same filter. It was found after the
run that refocusing the coronagraph between taking the flat fields and
the data meant that the flat field division left significant features
in the data, both at the edges of the coronagraph field and also
throughout the data at the location of what are presumed to be dust
particles on the coronagraph optics or CCD window. This problem was
particularly noticeable for the data taken on the last night of the
run.  However, the residuals are greatly reduced in the combined data.

Continuum sources were removed from the narrow-band images by
subtracting off a scaled version of the two 25{\AA} filter
observations. We investigated scaling methods, including measuring
stars or galaxies in the images, but found that the scaling derived was
consistent with simply subtracting off the average of the 25{\AA} filters
from each 13.5{\AA} observation.  That is, there was no evidence for a
significant continuum slope or calibration difference across the
150{\AA} region of interest, and the factor 2 shorter broad-band
exposure fairly accurately balanced the higher throughput of the
25{\AA} filters. This apparent consistency may be fortuitous: due to
the small field (small number of galaxies), and the errors in measuring
the flux of faint galaxies.  Figure~\ref{fig-coron} shows the sum of
the 25{\AA} data, and the continuum subtracted 13.5{\AA} data for
3C273. As can be seen, the PSF is a function of angle off-axis, and
becomes quite broad and asymmetric at the edge of the field, due to
aberrations in the coronagraph optics.

\subsection{Wide Field Imaging with COSMIC} \label{obscosmic}

A mosaic of Gunn r band images of the field around 3C273 was obtained
on 1992 February 23, at the Palomar 5m, with the prime focus COSMIC
system. This recently commissioned camera contains a thinned
2048$\times$2048 Tektronics CCD. Due to poor seeing, this was read out
with 2$\times$2 binning giving a scale of 0.56\arcsec/pixel. Exposures
were taken roughly centered on 4 positions offset from 3C273 by
5\farcm4 NE, NW, SE and SW. Four exposures of 5 minutes each were
obtained at each position. During the observations, the moon rose, and
so the background level of the images varies by almost a factor 2.

The data were reduced using IRAF. After bias subtraction, the data were
flat fielded using images of the dome. This left significant large
scale structure in the images, and so a `skyflat' was constructed from
the combined data images. This flat was median smoothed to remove any
small scale structure. Because of a region of very low sensitivity near
the center of the CCD, and also the location of a very bright star
coincidentally at the same place in one set of images, the flat fielded
data still shows a weak negative feature near the center of each
image.  The sky background was determined for each image separately and
subtracted. The offsets between the images were determined by measuring
the positions of the QSO and 2 bright nearby stars, which were common
to all images and the data were then mosaiced together, giving a
combined image with diameter 17\farcm2 centered on 3C273. The resulting
image is show as figure~\ref{fig-cosmic}.

\subsection{Fiber Spectroscopy} \label{obsfiber}

During 1992 February 8-10, spectra of objects in a 2.2$\times$1.6{\deg}
rectangle surrounding 3C273 were obtained with the Fiber Spectrograph
at the Las Campanas 2.5m  duPont Telescope. (\cite{she92}). This system
has 128 fibers which are manually plugged into an aluminum plate over a
1.5$\times$1.5{\deg} field. The fibers have a projected diameter of
3\farcs5. They feed the slit of a floor-mounted spectrograph. A
600 l/mm grating was used, with the 2D-Frutti photon-counting detector, giving
a
resolution of 8.6{\AA} FWHM. Three fields were observed for about 6000
seconds each, offset EW from each other.

Objects observed were chosen from a database produced by scanning an
UKST IIIaJ plate of the region with the APM scanning machine at
Cambridge. The APM produces a catalog of all the objects on a plate,
with estimates of isophotal magnitudes, size and a `sigma'
parameter that measures how much the image parameters differ from those
of stars with comparable magnitude. We chose objects to observe from
this catalog with `sigma'$>$3.0. This includes many fairly compact
objects, and resulted in a rather high contamination by stars, but also
means that we found a number of compact galaxies that would otherwise
have been missed. For each 1.5$\times$1.5{\deg} fiber field, a
magnitude sorted list of candidate galaxies was produced. Due to
restrictions on the minimum fiber separation, 330 out of a possible 336
object fibers were used, with 16 fibers per field set on blank sky.

The fiber spectra were extracted and reduced using the IRAF {\it apextract}
package.  The spectra were first traced and extracted. Then
fiber-to-fiber throughput differences were corrected with a flat field
image, and the fibers were wavelength calibrated and rebinned to a
common linear wavelength scale. The wavelength calibration used an arc
spectrum to determine the non-linear relationship between wavelength
and pixel number, after which a zero point shift was measured for each
fiber using the strong sky lines in the data. Finally, the sky emission
was subtracted from each fiber using an unscaled template constructed from the
16
sky fibers in each frame.

Classification and radial velocity measurements for each spectrum were
done in two ways. First each spectrum was inspected by eye, and
classified as either a star, galaxy or unknown. Then a radial velocity
was determined by measuring either the position of the 4000{\AA} break,
or that of the [OII] $\lambda$3727.6 feature (actually a doublet, but
unresolved in our data). A subjective assessment was also made of the
reliability of the resulting radial velocity, dividing the sample into
`possible' and `definite'. All the spectra were also analysed using the
{\it fxcor} routine in IRAF. Each object was cross-correlated with
three different templates: (a) a template made up from the 27 best S/N
stars in the data, (b) a template made up from 3 high S/N late-type
stars, and (c) a template made up from 16 emission line galaxies (all
shifted to their rest frames) in the data.  For each spectrum, the
correlation with the highest peak was then selected. A reassuringly
close match was found between the by-eye classification and velocities
and the results from {\it fxcor}. The resulting histogram of velocities
was inspected. Apart from 4 low S/N or hot stars (for which a good
template was not constructed), the stellar velocities found by {\it
fxcor} had a distribution well represented by a gaussian with zero mean
and an approximate dispersion ($\sigma$) of 85 km/s. We take this to be
a reasonable estimate for our radial velocity uncertainties.  It was
also found that the subjective `possible' category in the by-eye radial
velocity measurements matched rather well to a cross-correlation peak
height less than 0.3 returned by {\it fxcor}. Apart from one outlier,
the difference between the {\it fxcor} velocity and that measured by
eye for the galaxy identifications with cross-correlation peak heights
above 0.3 also was fairly well fit by a gaussian with $\sigma$ of 85
km/s.

In the end we obtained 129 definite galaxies, 43 possible galaxies, 86
definite stars, 4 possible stars, 10 fibers that had to be unplugged
due to over-illumination (which were hence almost certainly stars), 37
fibers which showed no flux (either very low surface brightness
galaxies or positional errors) and 21 spectra which showed flux but
which were unclassifiable either as stars or galaxies (based on very
low cross-correlation peak heights from fxcor and visual inspection).
We should reiterate that a much higher `success' rate in finding
galaxies could have been achieved by raising the cutoff `sigma' value
used in choosing galaxies from the APM scans, at the cost of missing
some compact galaxies.

We present in table~\ref{tbl-galcat} the resulting galaxy redshifts.
For each object, the table lists the RA and Dec, an approximate B
magnitude and the heliocentric radial velocity. The magnitudes were
calculated using the APM isophotal magnitudes measured from the plates,
crudely calibrated using the B magnitudes listed by \cite{sto80} and
\cite{sal92} for objects in the scanned region. They could be in error
by as much as 0.5.

A number of the brightest galaxies in the field were not included in
the fiber survey. Those with known redshifts within the survey region
were added to the sample (4 objects taken from the May 5 1990 version
of the CfA Redshift Catalog, \cite{huc90}). This gives a total sample
of 176 galaxies with redshifts  within the survey region which is
roughly complete to a B magnitude of 19.0.

We plot the results of the survey in a number of projections.
Figure~\ref{fig-zhist} shows a redshift histogram of the galaxies,
figure~\ref{fig-fiber-sky} shows the distribution of galaxies on the
sky, while figure~\ref{fig-pie} shows the pie-diagrams obtained.

\section{Analysis} \label{analysis}

In this section we go through each of the three observations described
in \S~\ref{observations} in turn, deriving constraints on the
absorbers. In \S~\ref{analline} we derive limits on H$\alpha$ line
emission at the velocities of the narrow-band filters and in
\S~\ref{analdwarf} we calculate the maximum absolute magnitude a dwarf
galaxy could have and remain undetected in our broad-band imaging.
A long description of the correlation analysis between the Ly$\alpha$
absorbers and the galaxies found in the fiber survey is given in
\S~\ref{analcorr}, in which the various available absorber and galaxy
sub-samples are discussed, and two alternative extreme hypotheses for
the absorber-galaxy correlation function are tested. Finally in
\S~\ref{particular} we discuss some particular aspects of the
absorber-galaxy distribution found in the 3C273 sightline.

\subsection{Flux Limits for H$\alpha$ Line Emission} \label{analline}

The final continuum-subtracted coronagraph images have a measured rms
dispersion (away from residuals due to bright stars) of 0.0018
DN/pixel/s. From calibration observations of HD84937, this is
equivalent to a 1$\sigma$ flux limit of 2.10$^{-18}$
ergs/cm$^2$/s/{\sq\arcsec}, equivalent to an emission measure of
approximately 2.8 pc cm$^{-6}$. By blinking the images, it can be seen
that none of the objects visible in the broad-band images have emission
line fluxes $>3\sigma$. For comparison, the VN2 image of Mkn49 showed a
peak H$\alpha$ line flux of 3.10$^{-14}$ ergs/cm$^2$/s/{\sq\arcsec},
i.e. 5,000 times higher than our 3$\sigma$ detection limit. One can
also perform the following thought experiment.  What would the Orion
nebula look like if placed at the distance of the Virgo cluster? As
discussed in \cite{ken84}, the Orion nebula is actually a relatively
low luminosity H~II region, with an H$\alpha$ luminosity of only
10$^{37}$ ergs/s.  Nevertheless, if placed at a distance of 16 Mpc, it
would still have a flux of 1.3 10$^{-15}$ ergs/cm$^2$/s (and would be
unresolved - the nebula has a diameter of 5 pc, while at Virgo,
1{\arcsec} corresponds to approximately 80 pc).  Thus it would be a
factor 220 brighter than our 3$\sigma$ limit. One can also calculate that the
Stromgren sphere around a single main sequence star of spectral class
$\sim$B1/{\sq\arcsec} would be detectable at the 3$\sigma$ level
(\cite{all73}).

Unfortunately, the expected surface brightness of an optically thick
slab of hydrogen, simply bathed in the local UV background would not be
detectable.  Taking the limit on the UV background from \cite{son89},
and using the formulae from \cite{ost89}, one finds a surface
brightness of $\le3.10^{-19}$ ergs/cm$^2$/s/{\sq\arcsec}, about a
factor 20 below our 3$\sigma$ limit.

Higher spectral and spatial resolution data have been taken by T.
Williams (private communication) using the Rutgers Fabry-Perot system
at the CTIO 4m telescope. This data has not been fully analysed, but
should produce even lower surface brightness limits (or detection) than our
data.

\subsection{Limiting Magnitude for detection of Low Surface Brightness Dwarf
Galaxies} \label{analdwarf}

The main motivation for taking the COSMIC images was to determine
whether there were any dwarf galaxies within a 40 kpc radius from the
line of sight to 3C273 which are too faint to be visible on the POSS
sky survey plates.  Examples of low surface brightness dwarfs in the
Virgo cluster can be seen in \cite{san84}. In order to determine how
faint a dwarf galaxy, which had morphological properties typical of
those found in the Virgo cluster, would be detectable in our image, we
used the IRAF {\it artdata} package to insert artificial dwarf galaxies
into the data array. The magnitude scale was derived by assuming that
the sky background in the Gunn r band (before the moon rose) was 21.5
(\cite{mas90}).
B absolute magnitudes were converted to Gunn r assuming a B-R color of
1.0 (B-V of 0.7), and a conversion from R to Gunn r of
r=R+0.43+0.15(B-V) (\cite{ken85}). Thus a dwarf galaxy with M$_{\rm
B}$=-15.5 was taken to have M$_{\rm r}$=-16.0. The Virgo distance
modulus was taken to be 31.02. Dwarf galaxy properties were taken from
\cite{bin85}. In particular, typical exponential scale
lengths\footnote{Scale Length=R$_0$, with Intensity $\propto{\rm
exp}(-1.6783\times{\rm R/R_{0}})$} for Virgo dwarfs were taken to be
2-4 kpc.  Figure~\ref{fig-artdata} shows the same data as
figure~\ref{fig-cosmic}, but with three dwarf galaxies added. One to
the NE with M$_{\rm B}$=-14.5 and scale length 4.4 kpc, one to the NW
with M$_{\rm B}$=-13.5 and scale length 2.2 kpc, and one to the SW with
M$_{\rm B}$=-13.5 and scale length 4.4 kpc. These experiments
demonstrated that we would be able to detect dwarf galaxies as faint as
-13.5 at the distance of Virgo, having morphological properties similar
to those found near the cluster center.  For comparison, the dwarf
galaxy illustrated in figure 2, panel 4, labelled `15{\deg}47' of
\cite{san84} has an absolute B magnitude of -14.8.

A complementary approach is to search for H~I 21 cm emission from Virgo
dwarf galaxies.  Recently, van Gorkom and her collaborators have set
extremely low limits with the VLA over a 40$\times$40 arcmin field
centered on 3C273 and over a velocity range of about 1000 km/s centered
on 1300 km/s to a 1 sigma column-density limit of approximately
10$^{19}$ cm$^{-2}$ (\cite{vang93a,vang93b}).

\subsection{Correlations between Lyman $\alpha$ Absorbers and Galaxies}
\label{analcorr}

We have shown there are no H~II regions, or other strong H$\alpha$
line-emitting gas,
or dwarf galaxies near the Virgo absorbers. Having
determined the redshift distribution of galaxies near the sightline to
3C273, (\S~\ref{obsfiber}) we would now like to address the statistical
question of the degree to which  the Lyman $\alpha$ absorbers are
correlated with luminous galaxies.  If they are correlated, is the Lyman
$\alpha$ absorber-galaxy correlation the same as the galaxy-galaxy
correlation? A cursory inspection of figure~\ref{fig-pie} is enough to
show that there will not be a simple answer to this question. While
there do seem to be Lyman $\alpha$ absorbers associated with clumps of
galaxies (e.g. the Virgo absorbers, or the set of absorbers around
z=0.02-0.03), there are also absorbers in conspicuous `voids' (at
z=0.06-0.07), and there are no absorbers associated with the prominent
excess of galaxies at z=0.078. We consider statistical tests for
various assumptions about the correlation between the Lyman $\alpha$
absorbers and the galaxies near the sightline towards 3C273 for which
we have redshift information. Ultimately, the goal should be a
quantitative and complete statistical description of the clustering
properties of the Lyman $\alpha$ absorbers themselves and their
correlation with various types of galaxies, clusters, voids etc. Given
both our rather meagre understanding of this problem and the small data
set, we shall concentrate on testing the following two extreme null
hypotheses about the Lyman $\alpha$ absorber clustering properties:

\begin{enumerate}
\item The Lyman $\alpha$ absorbers are uncorrelated with galaxies and are
randomly distributed. (The second part of this assumption necessarily
implies the first, but the converse is not necessarily true: The absorbers
could be correlated among themselves but be uncorrelated with galaxies)
\item The Lyman $\alpha$ absorbers are correlated with galaxies in the same
way that galaxies are correlated. More precise
formulations of this hypothesis depend upon the particular test
applied, as described below.
\end{enumerate}

In carrying out most of the tests described below, it is necessary to
compute the 3-dimensional distance between every absorber-galaxy or every
galaxy-galaxy pair. The question arises as to how to compute the
component along the line of sight, since departures from a perfectly
smooth Hubble flow distort the mapping of redshift onto radial
distance. With no information other than the angular coordinates
and redshifts for the objects, we cannot uniquely determine the
separation along the line of sight for any individual pair of objects.
For purposes of statistical tests we therefore make two different
assumptions about this component; the degree to which we do or do not
derive similar results will provide some indication of the sensitivity
of the test to the uncertainty in the estimation of this component:

\begin{enumerate}
\item We simply ignore any departures from Hubble flow
\item We adopt the formalism of \cite{dav83} to estimate this
component. These authors show how, knowing the two-point
correlation function for the {\it projected} distance between pairs,
in principal one can invert the integral equation relating the projected
distance
correlation function to the three-dimensional spatial
correlation function. However, for our limited and rather noisy sample,
this is not a very satisfactory procedure. Moreover, we would like
statistical estimates for the 3-dimensional separation for each pair
for the purposes of carrying out other types of tests (e.g. nearest
neighbor tests). We therefore use the Davis-Peebles formalism as
follows: Adopting the assumptions described in their paper, the
integrand in their equation (22) represents the probability that a pair
with projected separation r$_p$ Mpc and velocity difference $\pi$ km/s
has a separation in the radial direction of {\it y} Mpc.  We further
assume their functional form for {\it h(r)} in their equation (23),
with the parameter $F = 1$, and adopt their functional form for {\it
f(V)}, the probability of the relative velocity difference, as well as
their expression for $\sigma$, the dispersion in {\it f(V)}, i.e. their
equation (32). For $r_p >> r_o$ or $\pi >> H_o \times r_o $, where
$r_o$ is the characteristic correlation length, the probability is
strongly peaked at $ y \sim H_o \times \pi$, i.e., the pair separation
is very likely to be that given by assuming a pure Hubble flow.
However, as both of these inequalities fail to be satisfied, a second
maximum in the probability distribution arises at $y \approx 0$ whose
height depends upon the strength of the correlation - i.e., a pair with
a moderate velocity separation and small projected separation may be
separated by their Hubble flow distance, but if the correlation
between  such pairs is strong, this moderate velocity separation is
more likely to arise from a pair at about the same distance from us,
with strong gravitational interaction between them. As noted at the
outset of this discussion there is no way to unambiguously determine
the relative separation of any pair along the line of sight, but since
we are interested in statistical applications {\it we adopt the
expectation value of this probability distribution as our second
alternative algorithm}.
\end{enumerate}
A problem with this second approach is that evaluating the probability
distribution for the radial separation of the pair requires that we
know in advance the two-point correlation function between the pair.
Ideally, we could have dealt with this problem by using an iterative
approach: for both the absorber-galaxy and galaxy-galaxy pairs
separately, we start with some ``fiducial estimates'' for the two-point
correlation (e.g. $r_o=5.4 h^{-1}$ Mpc and $\gamma=1.77$ ; c.f.
equation 19 in Davis and Peebles) to compute the expectation value of
the radial separation for  each pair, and thus compute the two-point
correlation functions. With these new, separate best-fit values of
$\gamma$ and $r_o$ for the correlation functions for the
galaxy-absorber pairs and the galaxy-galaxy pairs, we could then repeat
the process until the parameters for the two-point correlation
functions have converged. In fact, since the results of our statistical
tests do not appear to be very sensitive to departures from a pure
Hubble flow, we have not carried out this iteration, but have simply
used the single set of parameters ($r_o=5.4 h^{-1}$ Mpc and
$\gamma=1.77$) in calculating the expectation values for both sets of
pairs.

In the following, we refer to these two algorithms for estimating the
radial separations  as the ``pure Hubble flow'' and ``perturbed Hubble
flow'' cases. When listing object separations in the following sections
we will give the separations found from the expectation value of the
perturbed Hubble flow model in brackets following the value for the
pure Hubble flow model.

Before carrying out any statistical tests we define the two samples and
discuss appropriate corrections for completeness.

\subsubsection{The Lyman $\alpha$ Sample} \label{lyasamp}

A carefully defined list of Lyman $\alpha$ absorption lines is
essential to a proper statistical discussion of the correlation
properties of Lyman $\alpha$ absorbers with galaxies. The preferred
list would obviously be drawn from a homogenous set of observations with the
smallest detectable equivalent width covering all or most of the
relevant redshift range.

Several line lists for the 3C273 sight line have been published
(\cite{mor91,bah91a,bah91b,bra93,bah93}). These line lists are compared
in table~\ref{tbl-lines}. However, not only are these lists based upon
3 different HST spectrograph configurations (GHRS G160M, GHRS G140L,
FOS G130) but they have also been produced by different reduction
procedures and line-finding and measuring algorithms and with differing
acceptance criteria for what constitutes a ``real'' line.  To
investigate the importance of this latter source of inhomogeneity we
have run the same continuum fitting, absorption line-finding and
line-measuring software, ``JASON'', used for the FOS line list, which
is described in detail by \cite{sch93}\footnote{We thank D. Schneider
for kindly making available the most recent version of JASON and for
instruction in its use.} on the data sets of \cite{mor91} and
\cite{bra93}. The JASON software was designed to run on FOS data with
an approximately gaussian point spread function (PSF). Unfortunately
this is not a good representation for the GHRS large aperture PSF, and
also in general the lines were resolved - meaning that the observed
line profiles had neither the instrumental PSF nor a gaussian shape.
The current version of JASON does not perform such convolutions, and so
we have run the search routines assuming a fixed PSF with the correct
(non-gaussian) shape for the GHRS, but with no account taken for
resolved lines. This means that the EWs output by the JASON software
are not accurate, but the detection significance levels are
approximately correct (see \cite{sch93}). The data set used is that
described by \cite{bra93}. We tabulate the significance levels from
JASON for the Lyman $\alpha$ lines in column 5 of
table~\ref{tbl-lines}, after the positions and EWs published in
\cite{bra93}. It can be seen that all of the `reliable' lines listed by
\cite{mor91} are confirmed by the JASON software, with the notable
exception of the line at $\lambda$1276.54, which was also not found by
\cite{bra93}. We checked this line by running the JASON software on the
original data used by \cite{mor91}, obtaining a significance level of
4.5 for the line. The line finding and fitting software used in
\cite{mor91} was developed by R. Carswell and J. Webb, and is described
in that paper. It used the GHRS PSF convolved with a Voit profile with
variable width.

Two other entries in table~\ref{tbl-lines} require special comment:  The line
at $\lambda$=1317.08 is identified as Lyman $\alpha$ in the list of
\cite{bah91b} whereas in the list of \cite{mor91} it is identified as
Ni~II.  This issue has been discussed in detail in \cite{bra93} who
give reasons for preferring the Ni~II identification which we adopt.
The second case involves the line at $\lambda$=1393.86. As discussed in
detail by \cite{sav93} this line appears to be a blend of one of the
members of the galactic Si~IV doublet and another strong line, whose
only plausible indentification is Lyman $\alpha$. The procedure used by
\cite{mor91} in estimating the strength of this line (which involves a
detailed comparison of the line profiles of the Si~IV doublet) is not
incorporated into the JASON formalism. For this reason we cannot assign
a formal uncertainty in the line strength. However the residuals from
an unblended fit to the Si~IV doublet are highly significant.

As a result of the above considerations, we have decided to adopt the
following samples of absorbers for our statistical tests:  (1) For the
Lyman $\alpha$ absorber ``{\it full} sample'' we adopt the list of 16
Lyman $\alpha$ absorbers (and their redshifts) given in \cite{mor91}
along with the additional low-redshift line ($\lambda$1224.52) given by
\cite{bah91a}; as noted in table~\ref{tbl-lines}, this last line is
visible in the GHRS G140L spectra but was below the significance
threshold of \cite{mor91}. This ``full sample'' is inhomogenous and/or
biased in three senses: i) The high resolution GHRS data covers only
redshifts above z$\sim$0.016; below this, only the FOS data of
\cite{bah91b} and the GHRS G140L low resolution data are available.
(Observations with the GHRS G160M grating in the redshift regime from
0.0 to 0.016 are scheduled for HST Cycle 3). Thus, the full sample may
be biased against weak low-redshift lines  in this redshift range that
may be detected by these Cycle 3 observations. ii) The line of sight
toward 3C273 may be somewhat atypical in that it passes through the
southern extension of the Virgo cluster. iii) Some of the weakest lines
listed as ``possible'' in the \cite{mor91} list may not be real.
Accordingly, along with the full sample of 17 lines, we shall also
consider two subsamples: (2) A ``{\it homogeneous} sample'' made of the
set of 14 lines found only with the GHRS G160M observations using the
original Carswell and Webb software (i.e. all but the first 3 lines in
column 1 of table~\ref{tbl-lines}). (As it happens, this is also
equivalent to deleting the 3 low-redshift (z$\la$0.016) lines possibly
associated with the Virgo cluster). (3) A ``{\it strong} sample''
composed of the set of 10 lines from \cite{mor91} with -log(P)$>$7.5
(but including the line at z=0.14658 for which a formal probability
estimate was not possible due to blending - i.e. the above sample with
the lines marked `d' in the comments column of table~\ref{tbl-lines}
removed). We have listed the sample membership in the final column of
table~\ref{tbl-lines}. Note that in contrast to the galaxy sample
discussed below there is no intrinsic observational selection against
the higher redshift absorbers.

One could, of course, define further samples. In particular, at the
request of the referee, we have also run our statistical tests of the
complete set of 5 lines listed in \cite{bah93} (i.e. the line list in
column 8 of table~\ref{tbl-lines}). Unfortunately, the number of lines
in this list is so small that neither of the two null hypotheses
considered below in connection with cloud-galaxy association can be
rejected with any significance. For this reason, and in order to keep
the various combinations of absorption line and galaxy samples to
managable proportions, we limit our tables of statistical results to
consideration of the three samples defined above.

\subsubsection{The Galaxy Sample} \label{galsamp}

An appropriate sample of galaxies with which to carry out the
correlation analysis would be one which is complete to some limiting
absolute magnitude throughout a cylindrical volume centered on the
3C273 sightline (i.e. out to a constant impact parameter) with a radius
large enough to sample most of the expected power in the correlation
function and length over (and beyond) the full redshift range covered
by the Lyman $\alpha$ line sample. The observed sample described in
\S~\ref{obsfiber} fails this requirement in two obvious respects: i) It
detects only the more luminous galaxies at the higher redshifts, ii) It
contains no galaxies with large impact parameters at low-redshifts. For
some tests these deficiencies are probably not important but for others
they are.  Accordingly, we will consider two galaxy samples. The first
is the sample described in \S~\ref{obsfiber}, which we will refer to as
the `cone' sample. The second is the cone sample together with all
galaxies from the May 5 1990 version of the CfA Redshift Catalog,
(\cite{huc90}) within 10 Mpc of the 3C273 line-of-sight. This gives a
heterogeneous sample with an unknown selection function, but is closer
to the ideal `filled cylinder' than the cone sample. It contains 1498
galaxies, the vast majority being at Virgo distances, and will be
referred to as the `cylinder' sample.

Thus for each hypothesis tested below there are 6 galaxy-absorber
sample combinations, and 2 possible estimates for the distance between
every pair of objects  giving a total of 12 data sets for each
statistical test.

\subsubsection{Tests of the 1st Null Hypothesis: The Lyman $\alpha$ absorbers
are uncorrelated with galaxies} \label{uncorr}

It is obvious that this hypothesis cannot literally be true - every
sightline that passes close to any galaxy, except those utterly devoid
of gas, will surely produce a detectable Lyman $\alpha$ line, and indeed
examples of this are already known (\cite{bah92a,bah92b}).
Nevertheless, in light of the fact that the high-redshift Lyman
$\alpha$ absorbers show almost no power in their two-point correlation
function (c.f. \cite{rau92} and references therein) it is of interest
to see if the present data set does or does not exclude this
hypothesis, and if it does, how strongly.

Having formulated this null hypothesis we consider two statistics as
measures of correlation (or lack of it):
\begin{enumerate}
\item The average over all the Lyman $\alpha$ absorbers of the distance
between a given Lyman $\alpha$ absorber and the N nearest galaxies in
the sample, with N=1,3 and 5.
\item The total number of absorber-galaxy pairs within a fixed radius R,
with R=500 kpc and 10 Mpc.
\end{enumerate}

In order to see whether the values of these observed statistics are
such that the null hypothesis can be rejected, we must find the
distribution of these same statistics for many realizations of the null
hypothesis. To do this, we carry out 1000 Monte Carlo simulations in
which the same number of absorbers as that of the particular Lyman $\alpha$
sample under consideration are laid down randomly (but follow the
``global'' distribution $dN/dz\sim(1+z)^{0.3}$ as determined by
\cite{bah93}). Redshift limits for the random absorbers were
0.03$<$z$<$0.151 for the total sample, and 0.016$<$z$<$0.151 for the
homogeneous and strong samples.

The results for the nearest-neighbour tests are summarised in
table~\ref{tbl-nearest}. We show results for the single nearest
neighbour galaxy, the mean of the 3 nearest and the mean of the 5
nearest.  For each case, the three columns list the observed mean
nearest-neighbour(s) distance, the average of the mean
nearest-neighbour(s) distances produced by the Monte Carlo simulations,
and the number of the 1000 Monte Carlo simulations that had a mean
nearest-neighbour distance less than the observed one. Thus this last
column, divided by 1000, can be taken as the probability that the
observed value could arise from a sample of absorbers distributed at
random with respect to the galaxies.

The most striking result is that there is a less than 0.1\% probability
that the average nearest-neighbour distance to the single closest
galaxy could arise from a randomly distributed set of absorbers. This
is true for all sample combinations,  and for
either pure or perturbed Hubble flow. For all samples, as one includes more
galaxies in the nearest-neighbour average, the significance drops.
After seeing this result, we wanted to test whether all the
significance came from the nearest galaxy, and so ran tests on the
second nearest galaxy only (also given in table~\ref{tbl-nearest}). One
can see that the observed mean distance is still significantly lower
than that expected for a randomly distributed set of absorbers, but
this could be explained by a combination of the highly significant
correlation with the nearest galaxy and the strong galaxy-galaxy
two-point correlation. This point is discussed in some detail by
\cite{phi93} in the context of bright galaxies found near quasar MgII
absorbers.

We also give in table~\ref{tbl-neargal} the RA, Dec and redshift of the
nearest galaxy in the `cylinder' sample to each Ly$\alpha$ absorbers of
the `total' sample (assuming pure Hubble flow).  The final column in
this table is the minimum absolute magnitude that could have been
detected in the fiber survey. This shows that for one absorber there is
no known galaxy with absolute magnitude above -17.8 within nearly 10
Mpc, (the same distance for the perturbed Hubble flow model) and that
even the nearest absorber-galaxy pair in our sample are separated by
350 kpc (240 kpc).  We will return to this in \S~\ref{particular}

The results for the number of galaxies within a fixed radius of each absorber
are given in table~\ref{tbl-ranrad}. These tests are essentially a
comparison of the integrated two-point correlation function out to the
given radius (see \cite{mo92b} for a discussion of this point). The
numbers given in the table are the observed number of absorber-galaxy
pairs within the given radius, the average of the number of pairs found
in 1000 Monte Carlo simulations with randomly distributed absorbers, and
the number of the Monte Carlo simulations with a larger number of pairs
than that observed. Thus the final column divided by 1000 is the
probability that the observed numbers of pairs or more would arise from
a random distribution of absorbers.

Table~\ref{tbl-ranrad} shows that there is no significant excess of
galaxies within volumes of radius 10 Mpc centered on the absorbers
compared to a random distribution, apart from the cylinder/total
subsample. For this combination there are a large number
of pairs between the Virgo absorbers and the many Virgo galaxies in
the CfA catalog. This result may be interpreted as saying that it is
surprising to find 3  out of 17 absorbers below z=0.008 (although see
\S~\ref{tgs}).  There is a marginally significant excess of
absorber-galaxy pairs within 500 kpc, over that expected for a random
distribution of absorbers, although the inclusion of the Virgo velocity
range removes the significance of this result. In summary, these tests
seem to be consistent with the nearest-neighbour distance results,
showing that there is an excess of close pairs of absorbers and
galaxies, but that this result vanishes if the averaging is done over
several galaxies or large radii.

\subsubsection{Tests of the 2nd Null Hypothesis: Identical Lyman $\alpha$
absorber-Galaxy, Galaxy-Galaxy Correlations} \label{ident-corr}

In order to test this hypothesis, one would like to use the same tests
as were used in \S~\ref{uncorr} A difficulty arises though in
generating a large number of Monte Carlo  samples.  We are loathe to
compare the observed distributions with simulations involving anything
other than the actual observed {\it galaxy} distribution since
differences between the observations and simulations (based, for
example on n-body or other galaxy clustering models) may result simply
from inadequacies of such models, and it is not clear how to create
simulations of (fake) absorber-(real) galaxy distributions having cross
correlation properties which are the same as the observed galaxy-galaxy
correlation properties.

One way to deal with this difficulty is to use a test
which does  not require the generation of Monte Carlo samples: If the
absorbers are distributed in the same way as the galaxies, then, given
the pencil-beam nature of the galaxy sample, the redshift distributions
of the absorbers and the galaxies should be identical, after correction
for differing selection effects in the two samples.  Table~\ref{tbl-ks}
shows the Kolmogorov-Smirnov D-values and probabilities that the
absorber redshift distribution is the same as that of the galaxies,
after the galaxy distribution is corrected using the selection function
shown in figure~\ref{fig-zhist}. This selection function was derived by
assuming a Schechter luminosity function with M$_*$=-19.5 and
$\alpha$=-0.97 (\cite{lov92}). Because this test requires a known
selection function, it can only be run for the `cone' galaxy sample.
Also, as it directly compares the redshifts, it does not require any
assumptions about pure or perturbed Hubble flow. It can be seen from
table~\ref{tbl-ks} that when all the absorption lines are included
there is a highly significant difference in the redshift
distributions.  This significance level becomes marginal when the Virgo
absorbers are removed, and vanishes when only strong absorbers beyond
Virgo are considered.

Our other test of the hypothesis that the absorbers and galaxies have
identical correlation functions uses the observed galaxy sample to
generate our Monte Carlo `absorber' sample. For each realization a
number of the actual galaxy redshifts were selected at random, and were
treated as absorbers on the 3C273 sightline. The actual algorithm
involved selecting at random a number of the observed galaxy redshifts
equal to the number in the absorber sample (making no correction for the galaxy
selection function, and with no restriction on how close together the
chosen galaxies were), and treating these redshifts as if they were
measured absorber redshifts on the 3C273 line of sight. The galaxies
which provided these redshifts were removed from the galaxy sample for
each test, to avoid an excess of `spurious' pairs\footnote{In fact
leaving in the galaxies would allow one to model a situation where the
absorbers were actually part of the halo of one of the observed
galaxies, although some maximum halo size would also have to be imposed
to make the simulation realistic. We are investigating this point in
more detail, and a paper is in preparation (\cite{mo93})}. The number of
absorber-galaxy pairs within a given radius were determined for both
the real and 1000 Monte Carlo samples, in an identical manner to the
second test in \S~\ref{uncorr} above. This procedure should be valid as
long as the radius within which the pair counts are being made is
significantly larger than the typical distance between a galaxy and the
3C273 line-of-sight. Because of this, tests were not run for a 500 kpc
radius. The results from these tests are given in
table~\ref{tbl-galrad}. The columns are the same as for
table~\ref{tbl-ranrad}, except that the final column lists the number
of Monte Carlo runs with {\it fewer} absorber-galaxy pairs than the
real sample. Thus this number divided by 1000 is the probability that
the absorbers could have a correlation function as strong as that
between galaxies. As can be seen, the absorber-galaxy correlation
function averaged over 10 Mpc is significantly weaker than that between
galaxies.

In summary, the two tests above seem to show that the absorber-galaxy
correlation function is significantly weaker than the galaxy-galaxy
correlation function over large scales (10 Mpc). Even though there is
clear evidence for galaxy-absorber clustering, {\it there is a
significant difference between the strength with which the galaxies are
clustered with respect to each other and the strength with which the
Lyman $\alpha$ absorbers are associated with galaxies.}

Some of the results of the two preceding sections can be inferred
directly by inspection of the actual two-point correlation functions
themselves.  A logarithmically binned version of the correlation
functions for pure and perturbed Hubble flows are shown in
figure~\ref{fig-2pt}. The absorber-galaxy correlation function was
generated using the total Lyman $\alpha$ sample and the cone galaxy
sample. The correlation functions were normalised in the usual way
using random samples with the selection function shown in
figure~\ref{fig-zhist}.  The error bars were estimated using the
formulae in \cite{mo92a}.  As expected, the pure and perturbed Hubble
flow models agree fairly well for separations larger than one or two
Mpc; the perturbed Hubble flow model produces more very close pairs,
since even small velocity differences wipe out small separations for
the pure Hubble flow model.  In both cases, the absorber-galaxy
correlation is clearly weaker than the galaxy-galaxy correlation on
scales from about 1-10 Mpc.  However, although visual inspection of the
absorber-galaxy correlation function may suggest that the correlation
is significant out to about 10 Mpc, in fact the pair tests summarized
in table~\ref{tbl-ranrad} and the nearest neighbor tests summarized in
table~\ref{tbl-nearest} both indicate that a statistically significant
absorber-galaxy correlations can be detected in our data sets only over
volumes which are smaller than this.

\subsection{Some Particular Cases of Interest} \label{particular}

\subsubsection{The Closest Absorber-Galaxy Associations} \label{closest}

Actual associations between individual observed galaxies and absorbers
in the 3C273 sight line are difficult to prove for a number of
reasons:  (1) The smallest projected distance to the 3C273
line-of-sight of all the galaxies in our sample is still 160 kpc.
(2) Galaxy rotation or velocity dispersions could produce velocity
differences as large as 200 km/s between the mean galaxy velocity and
an actually associated absorber (also comparable to the 3$\sigma$ error
in our fiber data velocity measurements) (3) In regions of high galaxy
density along the line-of-sight, peculiar motions of the galaxies and
the absorbers in cluster potential wells may make the velocity-distance
relationship complex. This is especially true of the Virgo region, and
for the z=0.0034 and 0.0053 absorbers (which also lie on the
steep portion of the damping wings of the galactic Lyman $\alpha$,
and have only been observed at low resolution). With these caveats, it
can be seen from table~\ref{tbl-neargal} that the closest absorber-galaxy
pair has a separation of 350 kpc (240 kpc). Outside of the Virgo region
(where the velocity-distance relationship may be less complex, but also
where our galaxy sample goes to much less faint absolute magnitudes),
the smallest separation is 410 kpc (350 kpc).

The best published example of an association between a Lyman $\alpha$
absorber and a galaxy is given in \cite{bah92a}, where they find a
galaxy within 90 kpc of the sightline to H1821+643 which has a strong
Lyman $\alpha$ absorber (EW 950 m{\AA}) within 400 km/s. There are no
galaxies in our sample this close to the line of sight (or indeed any
absorption systems this strong).

One can (somewhat arbitrarily) break the absorbers in our sample into
two groups: (1) those with a galaxy within 1 Mpc of the line-of sight, and
with a velocity difference of less than 400 km/s (all of the Virgo
systems and 6 of the higher-redshift systems. Despite its entry in
table~\ref{tbl-neargal}, the absorber at z=0.02622 actually has 2
galaxies within 400 km/s and 1 Mpc projected separation. They do not
appear in the table, as their large velocity differences ($>$250 km/s)
make their separation large, assuming pure Hubble flow), and (2) the rest (8
systems). There is no significant difference in the EW distribution of
these two samples.

Our data can also be used to consider the question: What is the average
galaxy diameter within which one would see a neutral-hydrogen column
density of at least 10$^{13}$ cm$^{-2}$? In practice this is rather a
naive question, as the cross-section almost certainly depends on galaxy
luminosity and probably also morphology. One might also expect a patchy
distribution of neutral-hydrogen in the outer parts of galaxies leading
to a covering factor not equal to one. Nevertheless, ignoring these
complications, and excluding the Virgo velocity range
due to (a) the possibility of large peculiar velocities (b) a wish to
avoid the large range of intrinsic galaxy luminosities in the sample
and (c) because our absorber EW limit is higher in this region, we find
that of the 12 galaxies with projected separation to the line-of-sight
less than 1 Mpc, 8 show Lyman absorption systems within 400 km/s. A
velocity difference of up to about 400 km/s could possibly be
attributed to internal motions within a large galaxy, coupled with our
measuring error, or alternatively to a small group.  Having 8 or more
such matches would occur 0.6\% of the time if the absorbers were
randomly distributed in velocity space.  However, of these 8 galaxies,
two are associated with the absorber at z=0.02622, and three with the
absorber at z=0.2933. Clearly several distinct galaxies can not be
producing the same absorber, and so in fact only 5 of the 12 systems
within 1 Mpc of the line-of-sight to the quasar can be legitimately
associated with individual absorbers. Reducing the projected separation
to 500 kpc, one finds 3 galaxies, with velocity separations of 20, 120
and 0 km/s (see table~\ref{tbl-nearest}).
This may be interpreted as saying that the cross-section of galaxies
with luminosities greater than 1/10 of L$^*$ for log(NH)$>10^{13}$ is
between 0.5 and 1 Mpc, although by ``cross-section'' we do not mean that the
absorbers are neccesarily associated with the actual galaxy in
question.

Because of the 3 Virgo absorbers, all of the Virgo galaxies have an
absorber within 400 km/s. However, it is clearly unreasonable to
suggest that more than one galaxy is associated with a given absorber.
Thus of the 19 Virgo galaxies within 500 kpc of the line-of-sight, 16
must not be producing absorption. Of these 19, 2 have absolute
magnitudes brighter than 1/10 L$^*$, and so are consistent with the
above statement, despite the higher absorber EW limit, as long as only
one of the galaxies with luminosity $<$1/10 L$^*$ is producing
observed absorption.

\subsubsection{The Most Isolated Lyman $\alpha$ Absorbers} \label{isolated}

One of the stronger lines in our absorber sample is also the most isolated.
Any galaxy brighter than
-18 would be in our sample at the redshift of the isolated z=0.06655
absorber. However, the nearest such galaxy has a projected separation
of  4.75 Mpc and a velocity difference of 710 km/s, which corresponds
to a spatial separation of 10 Mpc for both pure and perturbed Hubble
flow models.  Indeed studying the galaxy distribution in
figure~\ref{fig-pie}, the three absorbers with 0.060$<$z$<$0.07 seem to
lie in a `void' in the galaxy distribution. While it is certainly
possible that these absorbers are associated with galaxies below our
absolute magnitude limit,  {\it these isolated absorbers are a clear
demonstration that one can not associate ALL low-redshift Lyman
$\alpha$ absorbers with luminous (L$^*$) galaxies.}

\subsubsection{The Absence of Lyman $\alpha$ Absorbers in the z $\sim$ 0.078
Galaxy Concentration} \label{tgs}

The most striking feature of the observed galaxy redshift distribution is a
concentration of galaxies centered at a redshift of z $\sim$ 0.078.
This excess is almost certainly associated with  a structure which
includes the galaxy cluster Abell 1564. This cluster has a tabulated
center at 12:32:25 +02:07:11 (1950), giving it an offset of 89{\arcmin}
SW of the 3C273 sightline.  This angle corresponds to a separation of
7.3 Mpc at the cluster distance. Abell 1564 is the closest Abell
cluster to the 3C273 sightline, but is only richness class 0.
Redshifts for two member galaxies are given in \cite{met89}, giving an
average cluster redshift of 0.0793. Selecting all galaxies in our fiber
sample with redshifts between 0.07 and 0.085 (54 galaxies), one derives
a mean redshift of 0.0781 with rms of 0.0021 (630 km/s). For
comparison, selecting galaxies between redshifts of 0.15 and 0.17 (the
cluster around 3C273, see figure~\ref{fig-zhist}), one gets 24 galaxies
with mean redshift 0.1581 and rms 0.0024 (720 km/s).

Despite the location of Abell 1564 to the SW, the galaxies in the
redshift slice 0.07-0.085 show a weak concentration to the NE of the fiber
field, although there are galaxies in all parts of the fiber area
surveyed. If they are really part of a structure including A1564, the
structure must have a size of at least 8 Mpc. However, the nearest
Lyman $\alpha$ absorber is 0.0095 from the peak in redshift space (or
2850 km/s, which is 4.5 times the rms from the mean of the
concentration) and is at a distance of 36 Mpc from the center of the
concentration. It thus seems unlikely that there are any observed Lyman
$\alpha$ absorbers in our line of sight which are physically associated
with this concentration.

We can test a slightly different form of the hypothesis considered in
\S~\ref{ident-corr}, (though we grant its application to the z $\sim$
0.078 galaxy concentration involves {\it post facto} statistics),
namely:  What is the probability that, if the Lyman $\alpha$ absorbers
are distributed in space ``in the same way'' as the galaxies in our
sample, we should find the observed number of Lyman $\alpha$ absorbers
(none, in our case). By ``in the same way'' we mean that:
\[ \left(\rho_{Ly\alpha}(\vec{r})\over\overline{\rho_{Ly\alpha}}\right)
=\left(\rho_{galaxy}(\vec{r})\over\overline{\rho_{galaxy}}\right) \]
and by ``in the vicinity of'' we shall mean within $\pm2.5\times{\rm
rms}$ of the redshift peak of the concentration. We consider that our
selection function can be meaningfully applied over the redshift range
from about z= 0.016 out to our adopted ``proximity cutoff'' at z=
0.151.  Carrying out the integration of our sample (i.e.  the galaxy
numbers weighted by the selection function) between $\pm2.5\times{\rm
rms}$ of the peak of the galaxy concentration and over the redshift
range from 0.016 to 0.151 we find that  {\it after correction for the
selection function}, about 0.33 of an absolute-magnitude-limited
cylindrical sample of galaxies should be found within 2.5$\sigma$ of
the velocity peak of this structure. Since the Lyman $\alpha$ absorbers
are not subject to such a selection function,  we can apply this
equation to all the  volume elements along the line of sight and,
assuming further that the effective cross-section of the Lyman $\alpha$
absorbers ($\sigma$) is constant over the redshift range 0.016-0.151,
one can write:
\[ n\times\sigma\times[(z_{peak}+2.5\times{\rm rms})-(z_{peak}-2.5\times{\rm
rms})]\]
\[ =0.33\times\overline{n\times\sigma}\times[0.151-0.016] \]
This equation
represents the number of Lyman $\alpha$ absorbers in the vicinity of
the peak that we should expect to see if our hypothesis is correct.
Over the redshift interval 0.016-0.151  we have observed up to 14 Lyman
$\alpha$ absorbers so that we expect $ 0.33\times14\approx 4.7$
absorbers, and we observe none.  The probability of this occuring is
thus $exp(-4.7)\approx0.01$. This is not a strong result due to the
rather small number of absorbers expected, and the {\it post facto}
nature of the test as we acknowledged above, but is additional
suggestive evidence that the Lyman $\alpha$ absorbers do not follow the
galaxy distribution and in particular that they may avoid strong
concentrations of galaxies.  It could be strengthened, of course, by
observing additional lines of sight through dense concentrations.

It would be nice to repeat the above calculation for the case of the
three absorbers found in the Virgo velocity region. However, there are
a number of reasons why any estimate of the probability of finding
three absorbers in this region with an apparent overdensity of galaxies
is highly uncertain. Firstly, our absorber sample is incomplete in this
velocity range, as described in \S~\ref{lyasamp} Secondly, the area of
sky surveyed is so small that the selection function correction for
this velocity range is very large. Taking our selection function at
face value, the 8 galaxies actually in our survey at Virgo velocities
imply that 60\% of an absolute-magnitude-limited cylindrical galaxy
sample from z=0 to 0.151 would be found in that velocity range. Our
hypothesis above would then predict that (neglecting the incompleteness
of the absorber sample) one would expect 0.6$\times$17$\approx$10
absorbers, when we only see 3.  Unfortunately, it is not possible to
place a meaningful uncertainty on this apparent under-density of
absorbers in the Virgo region, for the reasons listed above.

In summary, there is marginal evidence, both from a clump of galaxies
at z=0.078 and also from the Virgo velocity range, that Lyman $\alpha$
absorbers are less common in regions of high galaxy density.

\section{Summary and Discussion} \label{sumdisc}

We have assembled several types of observations in an attempt to find
objects with which the Lyman $\alpha$ absorbers along the line of sight
to 3C273 might be associated, and in order to carry out statistical
tests of galaxy-Lyman $\alpha$ absorber association. In particular, we
obtained narrow-band images centered on and off the expected position
of any H$\alpha$ emission which might be associated with three of the
low redshift Lyman $\alpha$ absorbers, and obtained deep broad-band
images of a 17{\arcmin}$\times$17{\arcmin} field centered on 3C273.
Both these searches were negative.  Our failure to identify any
broad-band or H$\alpha$ emission from plausible ``galaxy-like'' objects
a few tens of kpc from the 3C273 sight line at the approximate distance
of the Virgo cluster will be checked by more sensitive and extensive
searches for H$\alpha$ by T.  Williams (private communication) and 21
cm emission by \cite{vang93a} and \cite{vang93b}.

We have also obtained redshifts for a large number of galaxies in the
vicinity of the 3C273 sightline. Again, we find no unambiguous instance
of association of any of the Lyman $\alpha$ absorbers with individual
galaxies. We define a number of samples for both the Lyman $\alpha$
absorbers and the galaxies and estimate the 3-dimensional separation
between each galaxy-galaxy pair and each Lyman $\alpha$ absorber-galaxy
pair based upon two models for converting the observed redshift
difference between any pair into a radial separation, viz. i) the
assumption of a pure Hubble flow and ii) a statistical model of
``perturbed'' Hubble flow based upon work of \cite{dav83}. The resulting
data base is used to carry out statistical
tests to confirm or reject two null hypotheses about the association of
galaxies and Lyman $\alpha$ absorbers, namely: i) The Lyman $\alpha$
absorbers show no tendency to cluster around galaxies ii) The Lyman
$\alpha$ absorbers cluster around galaxies exactly as the galaxies cluster
about each other.  While neither of these two hypotheses can be
unambiguously rejected in the sense that every combination of samples
and flow hypotheses reject both of them at significant
levels, the evidence from these tests, and from the galaxy-galaxy and
Lyman $\alpha$ absorber-galaxy two-point correlations themselves, points
quite strongly to the conclusion that both hypotheses are false.

In particular, over length scales from about 1 to 10 Mpc there seems
little doubt that the Lyman $\alpha$ absorbers cluster around galaxies
less strongly than the galaxies themselves cluster.  This is born out
by an examination of a redshift interval centered at about z=0.078 at
which a strong concentration of galaxies occurs but in the neighborhood
of which there are no Lyman $\alpha$ absorbers.  Additionally, we find
at least one Lyman $\alpha$ absorber for which no galaxy with absolute
magnitude brighter than about -18 can be found closer than about 5
Mpc.  Taken together, all this evidence suggests that the most
significant conclusion we have reached is that {\it the majority of
low-redshift Lyman $\alpha$ absorbers are not intimately associated
with normal luminous galaxies.}

In view of the fact that it has long been realized that at high
redshifts there is an  absence of power in the  Lyman $\alpha$ absorber
two-point correlation function in redshift space, except possibly at
the very smallest velocity separations, this conclusion is not too
surprising. On the other hand, it is also fairly clear that there is
{\it some} tendency for the Lyman $\alpha$ absorbers to cluster around
galaxies, and even weak evidence that this clustering becomes strong at
very small separations. Also \cite{bah92a} have investigated the
auto-correlation function of the low-redshift absorbers seen in the
line-of-sight to H1821+643, and show that there is only a 4\%
probability that the observed `clumping' arose from a randomly
distributed sample.

None of the foregoing points unambiguously, in our estimation, to a
particular model for the formation and evolution of the Lyman $\alpha$
absorbers. As have others, we simply offer the following speculations
which appear to be compatible with the facts as they are presently
understood:

At high redshifts, the Lyman $\alpha$ absorbers consist primarily of
entities which are only very loosely associated with larger mass
objects (e.g. proto galaxies), and which are evolving fairly rapidly.
Possibly this dominant population consists of absorbers in which
gravitational binding  (eg by dark matter) plays no significant role.
In addition to this group, there is a smaller population of absorbers
which are evolving less rapidly, possibly stabilized by dark matter
and which are clustered more strongly about galaxies. At very low
redshifts, this latter population is beginning to constitute a large
enough fraction of the absorbers that power in the two-point correlation
function is detectable. Thus, the present mix of Lyman $\alpha$
absorbers appears to have clustering properties intermediate between
present-epoch normal galaxies and a random non-clustered population.
This property also appears to be shared by the low-luminosity moderate
redshift ``blue galaxies'' (\cite{pri92}), leading to the plausible conjecture
that the
Lyman $\alpha$ absorbers are more closely related to low mass, low
luminosity galaxies than they are to L$^*$ galaxies, though the relation
is clearly not one-to-one.

We have no good way at present of estimating the characteristic scale
or masses of the low-redshift Lyman $\alpha$ absorbers. A guess at a
diameter of 30 kpc is as plausible as any. In particular, consider a
pancake whose diameter is 30 kpc and whose thickness is 10 kpc. In this
case, at the present epoch, a hydrogen column-density of
10$^{13}$-10$^{14}$ cm$^{-2}$ normal to the face of the absorber,
coupled with estimates for the present-epoch energy-density of
ionizating radiation leads to a total gas mass of order 10$^7$
M$_\odot$, but a mass of only a few hundred solar masses of {\it
neutral} hydrogen. It is of interest that the mass function of
neutral-hydrogen gas clouds appears to be truncated below about 10$^8$
M$_\odot$ (\cite{wei91}).  As Maloney (1992,1993) has shown, the gas in
a flaring galaxy with decreasing column-density will undergo a rather
sudden transition along its face from being mostly neutral to mostly
ionized, with the consequence that few if any contours with
neutral-hydrogen column-density of order 10$^{18}$ cm$^{-2}$ are
known.  Similarly, it is conceivable that, given the appropriate run of
length scale with mass, a sequence of masses would have the property of
making a sudden transition in the mass function of neutral-hydrogen
starting at about 10$^{8}$ M$_\odot$, leading to a dearth of objects
with total HI masses for several orders of magnitude below this.

If these speculations have any connection with reality then one might
expect to see some similarity in the clustering properties of the Lyman
$\alpha$ absorbers and low mass galaxies.  We are currently attempting
to obtain redshifts of galaxies of lower luminosity along the 3C273
sightline in order to investigate this possibility.

\acknowledgments

We would like to thank Greg Aldring and Steve Shectman for help and
advice in using the LCO fiber system, Neil Reid for giving us access to
one of the new Palomar Sky-survey plates, Alan Sandage for allowing us
to inspect one of his older IIIaJ plates, Mark Davis and Margaret
Geller for advice on generating Monte Carlo samples with known
correlation functions, Don Schneider for help with the JASON software
and Xavier Barcons, Donald Lynden-Bell, Mario Mateo, Houjon Mo, Michael
Rauch, John Webb, Jeff Willick and Dennis Zaritzky for useful
conversations.  We would also like to thank the referee, J. N. Bahcall,
for his helpful and constructive comments which improved the content
and presentation of the paper. SLM and RJW acknowledge support through
NASA contract NAS5-30101 and NSF grant AST-9005117.

\clearpage

\begin{table*}
\caption{Definite Galaxy redshifts in 3C273 Field} \label{tbl-galcat}
\end{table*}

\clearpage

\

\clearpage

\

\clearpage

\begin{table*}
\caption{Comparison of 3C273 Extragalactic Absorption Line Lists}
\label{tbl-lines}
\end{table*}

\clearpage

\begin{table*}
\caption{Nearest Neighbour Monte-Carlo tests} \label{tbl-nearest}
\end{table*}

\begin{table*}
\caption{Nearest Galaxies to 3C273 Lyman $\alpha$ absorbers}
\label{tbl-neargal}
\end{table*}

\clearpage

\begin{table*}
\caption{Pair Count Monte-Carlo tests using random absorbers}
\label{tbl-ranrad}
\end{table*}

\begin{table*}
\caption{KS comparison of z-distribution} \label{tbl-ks}
\end{table*}

\clearpage

\begin{table*}
\caption{Pair Count Monte-Carlo tests using the galaxy distribution}
\label{tbl-galrad}
\end{table*}

\clearpage

\begin{figure}
\caption{Coronagraph Data of 3C273 field. (a) Sum of 25{\AA} Filters,
the horizontal bar in the lower left is 1 armin long, (b) Continuum
Subtracted VN1, (c) Continuum Subtracted VN2, (d) Continuum Subtracted
HN. See text in \S~\protect\ref{obscoron}}
\label{fig-coron}
\end{figure}

\begin{figure}
\caption{Mosaic of COSMIC Images in Gunn r of 3C273 field. See text in
\S~\protect\ref{obscosmic} Region shown in figure~\protect\ref{fig-coron} is
circled.} \label{fig-cosmic}
\end{figure}

\begin{figure}
\caption{Histogram of redshifts for Galaxies found in the 3C273
sightline. Also plotted are the locations of Lyman $\alpha$ absorbers
from table~\protect\ref{tbl-nearest} (where `strong' refers to lines
with EW$\ge$55 m{\AA}, and `weak' refers to the remainder), and the
(arbitrarily normalised) galaxy selection function for the fiber
survey. See text in \S~\protect\ref{obsfiber}} \label{fig-zhist}
\end{figure}

\begin{figure}
\caption{Locations on the sky of the galaxies with redshifts in the
field of 3C273. See text in \S~\protect\ref{obsfiber}} \label{fig-fiber-sky}
\end{figure}

\begin{figure}
\caption{Pie-diagrams for the Galaxies observed with the LCO fiber
system. Angles have been exagerated by a factor 15 to prevent
overcrowding of the symbols. Please note that this results in a highly
distorted plot with initially spherical structures (such as the 3C273
cluster of galaxies) appearing elongated transverse to the line of
sight. Note also that the star marking the position of 3C273, while readily
visible in the projection in RA, is partially obscured by a clump
of galaxies in the projection in Dec} \label{fig-pie}
\end{figure}

\begin{figure}
\caption{Mosaic of COSMIC Images in Gunn r of 3C273 field, with 3
artificial dwarf galaxies added. NE with M$_{\rm B}$=-14.5 and scale
length 15kpc, NW with M$_{\rm B}$=-13.5 and scale length 7.5kpc, and SW
with M$_{\rm B}$=-13.5 and scale length 15kpc. See text in
\S~\protect\ref{analdwarf}}
\label{fig-artdata}
\end{figure}

\begin{figure}
\caption{ Two point correlation functions for galaxy-galaxy and
absorber-galaxy (a) assuming pure Hubble flow (b) assuming perturbed Hubble
flow.  See text in
\S~\protect\ref{ident-corr}} \label{fig-2pt}
\end{figure}


\clearpage

\begin{thebibliography}{}
\bibitem[Allen 1973]{all73} \reference Allen, C. W. 1973, ``Astrophysical
Quantities 3rd Edition'', Athlone Press, London, Pg 267
\bibitem[Bahcall {\it et al.} 1991a]{bah91a} \reference Bahcall, J. N.,
Jannuzi, B. T., Schneider, D. P., Hartig, G. F., Bohlin, R. and Junkkarinen,
V., 1991a, in ``The First Year of HST Observations'', ed A. L. Kinney and J. C.
Blades, Pg 46
\bibitem[Bahcall {\it et al.} 1991b]{bah91b} \reference Bahcall, J. N.,
Jannuzi, B. T., Schneider, D. P., Hartig, G. F., Bohlin, R. and Junkkarinen,
V., 1991b, \apjl, 377, L5
\bibitem[Bahcall {\it et al.} 1992a]{bah92a} \reference \reference Bahcall, J.
N., Jannuzi, B. T., Schneider, D. P., Hartig, G. F. and Green, R. F., 1992a,
\apj, 397, 68
\bibitem[Bahcall {\it et al.} 1992b]{bah92b} \reference Bahcall, J. N.,
Jannuzi, B. T., Schneider, D. P., Hartig, G. F. and Jenkins, E. B., 1992b,
\apj, 398, 495
\bibitem[Bahcall {\it et al.} 1993]{bah93} \reference Bahcall, J. N., Bergeron,
J., Boksenberg, A., Hartig, G. F., Jannuzi, B. T., Kirhakos, S., Sargent, W. L.
W., Savage, B. D., Schneider, D. P., Turnshek, D. A., Weymann, R. J. and Wolfe,
A. M., 1993, \apjs, In Press
\bibitem[Bingelli, Sandage and Tammann 1985]{bin85} \reference Binggeli, B.,
Sandage, A. and Tammann, G. A., 1985, \aj, 90, 1681
\bibitem[Brandt {\it et al.} 1993]{bra93} \reference Brandt, J. C. {\it et
al.}, 1993, \aj, 105, 831
\bibitem[Chernomordic and Ozernoy 1983]{che83} \reference Chernomordic, V. V.
and Ozernoy, L. L., 1983, Nature, 303, 153
\bibitem[Davis and Peebles 1983]{dav83} \reference Davis, M. and Peebles, P. J.
E. 1983, \apj, 267, 465
\bibitem[Fransson and Epstein 1982]{fra82} \reference Fransson, C. and Epstein,
R., 1982, \mnras, 198, 1127
\bibitem[Haynes and Giovanelli 1984]{hay84} \reference Haynes, M.P. and
Giovanelli, R. 1984, \aj, 89, 758
\bibitem[Huchra 1990]{huc90} \reference Huchra, J. P. 1990 May 5 1990 version
of the CfA Redshift Catalog, obtained through the Astronomical Data Center.
\bibitem[Jacoby {\it et al.} 1992]{jac92} \reference Jacoby, G. H., Branch, D.,
Ciardullo, R., Davies, R. L., Harris, W. E., Pierce, M. J., Pritchet, C. T.,
Tonry, J. L. and Welch, D. L., 1992, \pasp, 104, 599
\bibitem[Kennicutt 1984]{ken84} \reference Kennicutt, R. C., 1984, \apj, 287,
116
\bibitem[Kent 1985]{ken85} \reference Kent, S. M., 1985, \pasp, 97, 165
\bibitem[Loveday {\it et al.} 1992]{lov92} \reference Loveday, J., Peterson, B.
A., Efstathiou, G. and Maddox, S. J., 1992, \apj, 390, 338
\bibitem[Maloney 1992]{mal92} \reference Maloney, P., 1992, \apjl, 389, L89
\bibitem[Maloney 1993]{mal93} \reference Maloney, P., 1993, \apj, In Press
\bibitem[Massey 1990]{mas90} \reference Massey, P., 1990, NOAO Newsletter \#
21, March 1990
\bibitem[Mo, Jing and Borner 1992]{mo92a} \reference Mo, H. J., Jing, Y. P. and
Borner, G., 1992, \apj, 392, 452
\bibitem[Mo {\it et al.} 1992]{mo92b} \reference Mo, H. J., Einasto, M., Xia,
X. Y. and Deng, Z. G., 1992, \mnras, 255, 382
\bibitem[Mo, Morris and Weymann 1993]{mo93} \reference Mo, H. J., Morris, S. L.
and Weymann, R. J., In Preparation
\bibitem[Metcalfe {\it et al.} 1989]{met89} \reference Metcalfe, N., Fong, R.,
Shanks, T. and Kilkenny, D. 1989, \mnras, 236, 207
\bibitem[Morris {\it et al.} 1991]{mor91} \reference Morris, S. L., Weymann, R.
J., Savage, B. D. and Gilliland, R. L., 1991, \apjl, 377, L21
\bibitem[Osterbrock 1989]{ost89} \reference Osterbrock, D. E., ``Astrophysics
of Gaseous Nebulae and Active Galactic Nuclei'', 1989, University Science
Books, CA
\bibitem[Phillips, Disney and Davies 1993]{phi93} \reference Phillips, S.,
Disney, M. J. and Davies, J. I., 1993, \mnras, 260, 453
\bibitem[Pritchet and Infante 1992]{pri92} \reference Pritchet, C. J. and
Infante, L., 1992, \apj, 399, L35
\bibitem[Rauch {\it et al.} 1992]{rau92} \reference Rauch, M, Carswell, R. F.,
Chaffee, F. H., Foltz, C. B., Webb, J. K., Weymann, R. J., Bechtold, J. and
Green, R. F., 1992, \apj, 390, 387
\bibitem[Salzer 1992]{sal92} \reference Salzer, J. J. 1992, \aj, 103, 385
\bibitem[Sandage and Bingelli 1984]{san84} \reference Sandage, A. and Bingelli,
B., 1984, \aj, 89, 919
\bibitem[Savage {\it et al.} 1993]{sav93} \reference Savage, B. D., Lu, L.,
Weymann, R. J, Morris, S. L. and Gilliland, R. L., 1993, \apj,  404, 124
\bibitem[Schneider {\it et al.} 1993]{sch93} \reference Schneider, D. P.,
Hartig, G. F., Jannuzi, B. T., Kirhakos, S., Saxe, D. H., Weymann, R. J.,
Bahcall, J. N., Bergeron, J., Boksenberg, A., Sargent, W. L. W., Savage, B. D.,
Turnshek, D. A. and Wolfe, A. M., 1993, \apjs, In Press
\bibitem[Shectman 1992]{she92} \reference Shectman, S. A., 1992, in ``Fiber
Optics in Astronomy II'', ed P. M. Gray, ASP Volume 37, Pg 26
\bibitem[Songaila, Bryant and Cowie 1989]{son89} \reference Songaila, A.,
Bryant, W. and Cowie, L. L., 1989, \apj, 345, L71
\bibitem[Stockton 1980]{sto80} \reference Stockton, A. 1980, In IAU symposium
92, ``Objects of High Redshift'', ed G. O. Abell and P. J. E. Peebles, Pg 87
\bibitem[van Gorkom 1993]{vang93a} \reference van Gorkom, J., H., 1993, Teton
Conference Proceedings, In Press, Ed M. Shull and H. Thronson
\bibitem[van Gorkom {\it et al.} 1993]{vang93b} \reference van Gorkom, J., H.,
Bahcall, J.N., Jannuzi, B., and Schneider.D. 1993, \aj, In Press
\bibitem[Vilas and Smith 1987]{vil87} \reference Vilas, F. and Smith, B. A.,
1987, \ao, 26, 4
\bibitem[Weinberg {\it et al.} 1991]{wei91} \reference Weinberg, D. H.,
Szomoru, A., Guhathakurta, P. and van Gorkom, J. H., 1991, \apjl, 372, L13
\end{thebibliography}
\end{document}